# $\gamma^* p \to \Delta$ FORM FACTORS IN QCD

V.M. Belyaev*

*Continuous Electron Beam Accelerator Facility*
*Newport News, VA 23606, USA*
E-mail: belyaev@cebaf.gov
and
*ITEP, 117259, Moscow, Russia*

and

A.V. Radyushkin

*Physics Department, Old Dominion University*
*Norfolk, VA 23529, USA*
and
*Continuous Electron Beam Accelerator Facility*
*Newport News, VA 23606, USA*

## ABSTRACT

We use local quark-hadron duality to estimate the purely nonperturbative soft contribution to the $\gamma^* p \to \Delta$ form factors. Our results are in agreement with existing experimental data. We predict that the ratio $G_E^*(Q^2)/G_M^*(Q^2)$ is small for all accessible $Q^2$, in contrast to the perturbative QCD expectations that $G_E^*(Q^2) \to -G_M^*(Q^2)$.

**1.**

There are two competing explanations of the experimentally observed power-law behaviour of hadronic form factors: hard scattering [1] and the Feynman mechanism [2]. At sufficiently large momentum transfer, the hard scattering mechanism dominates[3]. However, there is increasing evidence that, for experimentally accessible momentum transfers, the form factors are still dominated by the soft contribution corresponding to the Feynman mechanism [4].

In this talk, we consider the soft mechanism contribution to the $\gamma^* p \to \Delta$ form factors. The relevant hard scattering contribution was originally considered in ref. [5].

Here, we use the local quark-hadron duality to estimate the soft contribution for the $G_E^*(Q^2)$ and $G_M^*(Q^2)$ form factors of the $\gamma^* p \to \Delta$ transition.

The starting object for a QCD sum rule analysis [6] of the $\gamma^* p \to \Delta$ transition is the 3-point correlator:

$$T_{\mu\nu}(p,q) = \int \langle 0 | T\{\eta_\mu(x) J_\nu(y) \bar{\eta}(0)\} | 0 \rangle e^{ipx-iqy} d^4x d^4y \tag{1}$$

---

*Contribution to the International Symposium and Workshop on Particle Theory and Phenomenology at Iowa State University, Ames, May 22-24

of the electromagnetic current

$$J_\nu = e_u \bar{u} \gamma_\nu u + e_d \bar{d} \gamma_\nu d \tag{2}$$

and two Ioffe currents [7]

$$\eta = \varepsilon^{abc} \left( u^a \mathcal{C} \gamma_\rho u^b \right) \gamma_\rho \gamma_5 d^c \ , \quad \eta_\mu = \varepsilon^{abc} \left( 2 \left( u^a \mathcal{C} \gamma_\mu d^b \right) u^c + \left( u^a \mathcal{C} \gamma_\mu u^b \right) d^c \right) \ . \tag{3}$$

On the hadronic level, the contribution of $\gamma^* p \to \Delta$ transition to (1) is:

$$T^{\gamma^* p \to \Delta}_{\mu\nu} = \frac{l_N l_\Delta}{(2\pi)^4} \frac{X_{\mu\alpha}(p)}{p^2 - M^2} \Gamma_{\alpha\nu}(p,q) \gamma_5 \frac{\hat{p} - \hat{q} + m}{(p-q)^2 - m^2} \ , \tag{4}$$

where $\Gamma_{\alpha\nu}(p,q)\gamma_5$ is the $\gamma^* p \to \Delta$ vertex function

$$\Gamma_{\alpha\nu}(p,q) = G_1(q^2)(q_\alpha \gamma_\nu - g_{\alpha\nu} \hat{q}) + G_2(q^2)(q_\alpha P_\nu - g_{\alpha\nu}(qP)) \\ + G_3(q^2)\left(q_\alpha q_\nu - g_{\alpha\nu} q^2\right) \tag{5}$$

($P \equiv p - q/2$), $l_N$ and $l_\Delta$ are the residues of nucleon and $\Delta$ of the quark currents (3), $X_{\mu\alpha}(p)$ the projector onto the isobar state

$$X_{\mu\alpha}(p) = \left( g_{\mu\alpha} - \frac{1}{3}\gamma_\mu \gamma_\alpha + \frac{1}{3M}(p_\mu \gamma_\alpha - p_\alpha \gamma_\mu) - \frac{2}{3M^2} p_\mu p_\alpha \right)(\hat{p} + M). \tag{6}$$

The form factors $G_1, G_2, G_3$ are related to a more convenient set $G^*_E, G^*_M, G^*_C$ by

$$G^*_M(Q^2) = \frac{m}{3(M+m)} \left( ((3M+m)(M+m) + Q^2) \frac{G_1(Q^2)}{M} \right. \\ \left. + (M^2 - m^2) G_2(Q^2) - 2Q^2 G_3(Q^2) \right) \ , \tag{7}$$

$$G^*_E(Q^2) = \frac{m}{3(M+m)} \left( (M^2 - m^2 - Q^2) \frac{G_1(Q^2)}{M} \right. \\ \left. + (M^2 - m^2) G_2(Q^2) - 2Q^2 G_3(Q^2) \right) \ , \tag{8}$$

$$G^*_C(Q^2) = \frac{2m}{3(M+m)} \left( 2M G_1(Q^2) + \frac{1}{2}(3M^2 + m^2 + Q^2) G_2(Q^2) \right. \\ \left. + (M^2 - m^2 - Q^2) G_3(Q^2) \right) \ , \tag{9}$$

2.

To incorporate the local quark-hadron duality, we write down the dispersion relation for each of the invariant amplitudes:

$$T_i(p_1^2, p_2^2, Q^2) = \frac{1}{\pi^2} \int_0^\infty ds_1 \int_0^\infty ds_2 \frac{\rho_i(s_1, s_2, Q^2)}{(s_1 - p_1^2)(s_2 - p_2^2)} + \text{``subtractions''} \ , \tag{10}$$

where $p_1^2 = (p - q)^2$, $p_2^2 = p^2$. The perturbative contributions to the amplitudes $T_i(p_1^2, p_2^2, Q^2)$ can also be written in the form of eq.(10). The local quark-hadron duality assumes, that the two spectral densities are in fact dual to each other:

$$\int_0^{s_0} ds_1 \int_0^{S_0} \rho_i^{pert.}(s_1, s_2, Q^2) \, ds_2 = \int_0^{s_0} ds_1 \int_0^{S_0} \rho_i(s_1, s_2, Q^2) \, ds_2 \,, \qquad (11)$$

For the tensor structures of $T_{\mu\nu}$, it is convenient to use the basis in which $\gamma_\mu$ is placed at the leftmost position. Then, the invariant amplitudes corresponding to the structures with $q_\mu$ and $g_{\mu\nu}$ are free from the contributions due to the spin-1/2 isospin-3/2 states.

The number of independent amplitudes can be diminished by taking some explicit projection of the original amplitude $T_{\mu\nu}(p, q)$. In particular, the invariant amplitude corresponding to the structure $q_\mu[\hat{q}, \hat{p}]$ for the amplitude $T_{\mu\nu} p_\nu$ is proportional to the quadrupole form factor $G_C^*(Q^2)$:

Another possibility is to take the trace of $T_{\mu\nu}^{\gamma^* p \to \Delta}$. The result is proportional to the magnetic form factor $G_M^*(Q^2)$: However, one should remember that the trace of $T_{\mu\nu}$ is not free from contributions due to spin-1/2 isospin-3/2 states.

Though the invariant amplitude related to the trace of $T_{\mu\nu}$ is contaminated by the transitions into spin-1/2 isospin-3/2 states, it makes sense to consider this amplitude because it has the simplest perturbative spectral density:

$$\frac{1}{\pi^2} \rho_M^{pert.}(s_1, s_2, Q^2) = \frac{Q^2}{8\kappa^3} (\kappa - (s_1 + s_2 + Q^2))^2 (2\kappa + s_1 + s_2 + Q^2) \,, \qquad (12)$$

where

$$\kappa = \sqrt{(s_1 + s_2 + Q^2)^2 - 4 s_1 s_2} \,. \qquad (13)$$

Imposing the local duality prescription, we get

$$G_M^*(Q^2) = \frac{2m}{l_N l_\Delta (M + m)} \int_0^{s_0} ds_1 \int_0^{S_0} \frac{\rho_M^{pert.}(s_1, s_2, Q^2)}{\pi^2} ds_2$$

$$= \frac{6m}{(M + m)} F(s_0, S_0, Q^2) \,, \qquad (14)$$

where $F(s_0, S_0, Q^2)$ is a universal function

$$F(s_0, S_0, Q^2) = \frac{s_0^3 S_0^3}{9 l_N l_\Delta (Q^2 + s_0 + S_0)^3 \left(1 - 3\sigma + (1 - \sigma)\sqrt{1 - 4\sigma}\right)} \qquad (15)$$

and $\sigma = s_0 S_0 / (Q^2 + s_0 + S_0)^2$. We fix the nucleon duality interval $s_0$ at the standard value $s_0 = 2.3 \, GeV^2$ extracted from the analysis of the two-point function and used earlier in the nucleon form factor calculations. To fine-tune the $S_0$ value, we consider two independent sum rules for the $G_1$ form factor

$$mG_1(Q^2) = 2\left(3 + Q^2\frac{d}{dQ^2}\right) F(s_0, S_0, Q^2) - 2Q^2 \left(\frac{d}{dQ^2}\right)^2 \int_0^{S_0} F(s_0, s_2, Q^2)\, ds_2 \quad (16)$$

and

$$MG_1(Q^2) = \frac{3}{2}Q^2 \left(\frac{d}{dQ^2}\right)^2 \int_0^{S_0} F(s_0, s_2, Q^2)\, ds_2 \quad (17)$$

corresponding to the structures $q_\mu[\gamma_\nu, \hat{p}]$ and $q_\mu[\gamma_\nu, (\hat{p} - \hat{q})]$.

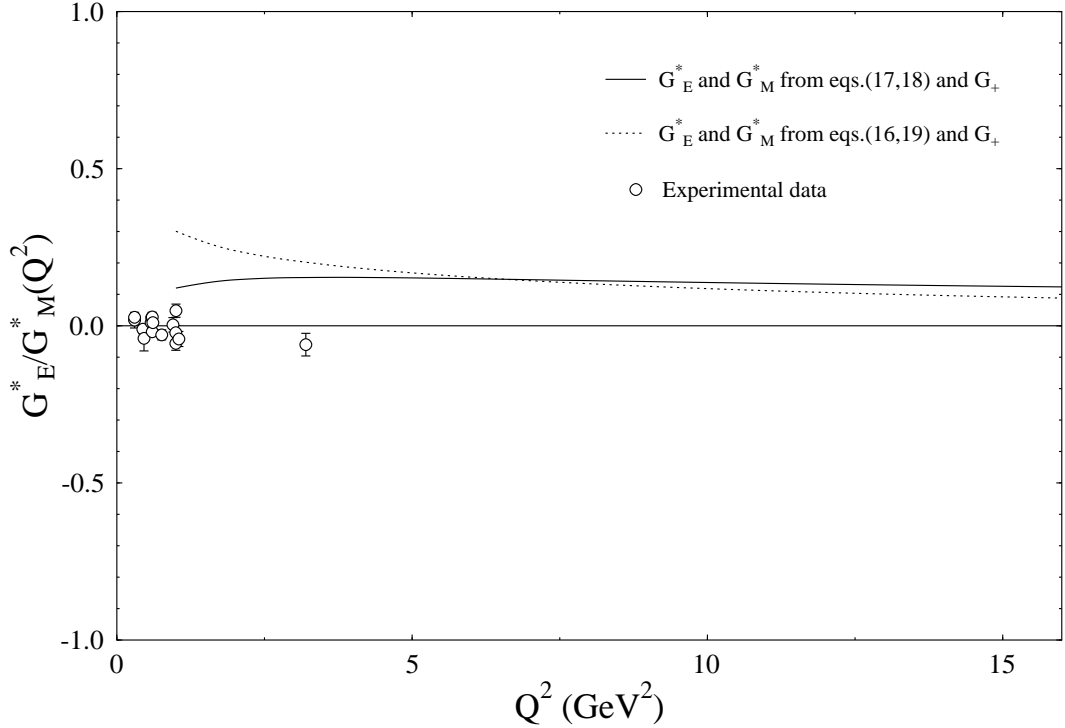

Fig. 1. Ratio of Form Factors: $G_E^*(Q^2)/G_M^*(Q^2)$

Taking the ratio of these two relations, one can investigate their mutual consistency and test the overall reliability of the quark-hadron duality estimates. The best agreement is reached for $S_0 = 3.5\, GeV^2$, and we will use this value as the basic isobar duality interval in further calculations.

From eqs.(7) and (8), it follows that $G_1$ is proportional to the difference of the magnetic $G_M^*$ and electric $G_E^*$ transition form factors:

$$G^{(-)}(Q^2) \equiv G_M^*(Q^2) - G_E^*(Q^2) = \frac{2m}{3M(M+m)} \left((M+m)^2 + Q^2\right) G_1(Q^2)\,. \quad (18)$$

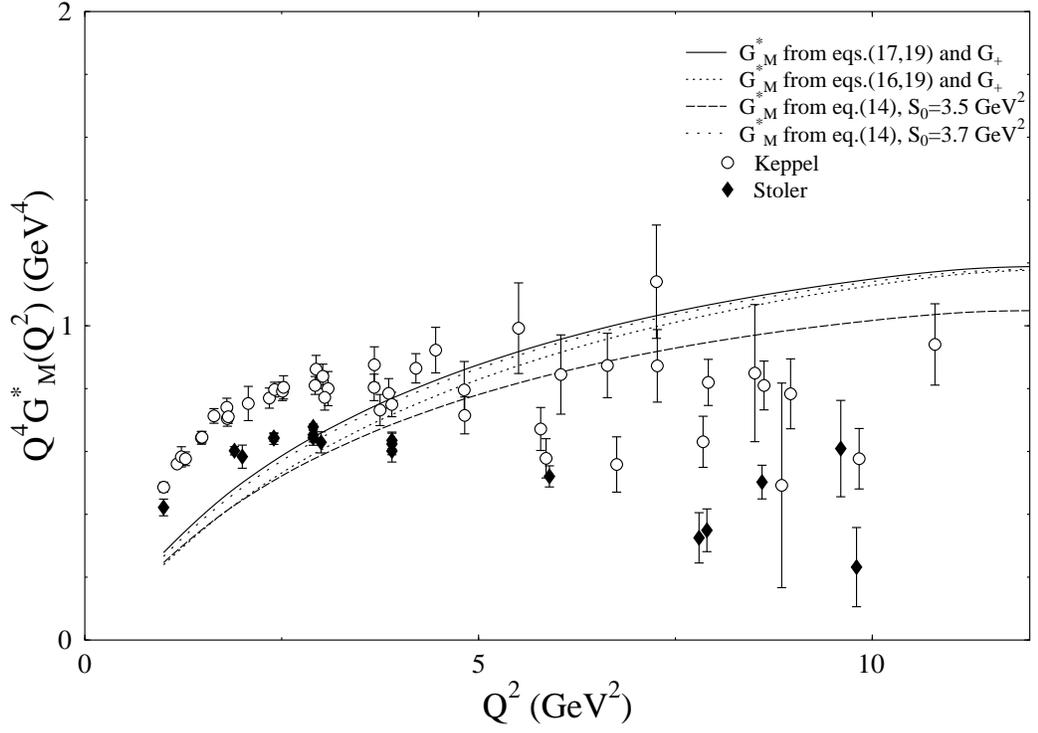

Fig. 2. Form Factor $G_M^*(Q^2)$

The sum $G^{(+)}(Q^2) \equiv G_M^*(Q^2) + G_E^*(Q^2)$ of these form factors can be obtained from the invariant amplitude corresponding to the structure $g_{\mu\nu}[\hat{p},\hat{q}]$:

$$G^{(+)}(Q^2) = \frac{8m}{M+m}\left[F(s_0, S_0, Q^2) - \frac{Q^2}{12}\left(\frac{d}{dQ^2}\right)^2 \int_0^{s_0} F(s_1, S_0, Q^2)ds_1\right] . \qquad (19)$$

An important observation is that $G_E^*(Q^2)$ is predicted to be much smaller than $G_M^*(Q^2)$ (see Fig.1). It should be noted that pQCD approach predicts, that $G_M^*(Q^2) \simeq -G_E^*(Q^2)$ for asymptotically large $Q^2$.

One should realize, that $G_E^*(Q^2)$ is obtained in our calculation as a small difference between two large combinations $G^{(+)}(Q^2)$ and $G^{(-)}(Q^2)$, both dominated by $G_M^*(Q^2)$, so we restrict ourselves to a conservative statement that the electric form factor $G_E^*(Q^2)$ is small compared to $G_M^*(Q^2)$.

Experimental points for $G_M^*$ shown in Fig.2 were taken from the results for the $G_T(Q^2)$ form factor obtained from analysis of inclusive data [8], [9], [10]. One can see that, in the $Q^2 \gtrsim 3\, GeV^2$ region, the local duality predictions $G_M^*(Q^2)$ are close to the results of the recent analysis by C. Keppel (see [10]).

The quadrupole (Coulomb) form factor $G_C^*(Q^2)$ has been calculated also. We obtained that $G_C^*(Q^2)$ is essentially smaller than $G_M^*(Q^2)$ and has an extra $1/Q^2$ suppression compared to $G_M^*(Q^2)$.

**3.**

We applied the local quark-hadron duality prescription to estimate the soft contribution to the $\gamma^* p \to \Delta$ transition form factors. We observed a reasonable agreement between the results obtained from different invariant amplitudes. We found that the transition is dominated by the magnetic form factor $G_M^*(Q^2)$ while electric $G_E^*(Q^2)$ and quadrupole $G_C^*(Q^2)$ form factors are small compared to $G_M^*(Q^2)$ for all experimentally accessible momentum transfers. Numerically, our estimates for $G_T(Q^2)$ are close to those obtained from a recent analysis of inclusive data [10]. Hence, there is no need for a sizable hard-scattering contribution to describe the data. Furthermore, if future exclusive measurements at CEBAF would show that the ratio $G_E^*(Q^2)/G_M^*(Q^2)$ is small above $Q^2 \sim 3\,GeV^2$, this would give an unambiguous experimental proof of the dominance of the soft contribution.

**4. Acknowledgements**

We are very grateful to P.Stoler, N.Isgur, V.Burkert and C.E.Carlson for discussions which strongly motivated this investigation. We thank C. Keppel for providing us with the results of her analysis [10].

This work was supported by the US Department of Energy under contract DE-AC05-84ER40150.